\documentclass{article}

\usepackage[fleqn]{amsmath}
\usepackage[mathscr]{eucal}
\usepackage{amsfonts}
\usepackage{amssymb}

\newcommand{\apriori}{\emph{a priori}}

\newcommand{\refeq}[1]{(\ref{#1})}

\newcommand{\vect}[1] {\boldsymbol{{ #1}} }

\newcommand{\Csp}{\mathfrak{C}}
\newcommand{\Hsp}{\mathfrak{H}}

\newcommand{\Lsp}{\mathfrak{L}}
\newcommand{\Msp}{\mathfrak{M}}
\newcommand{\Ssp}{\mathfrak{S}}

\newcommand{\Dset}{\mathbb{D}}

\newcommand{\Lset}{\mathbb{L}}
\newcommand{\Mset}{\mathbb{M}}
\newcommand{\Nset}{\mathbb{N}}

\newcommand{\Rset}{\mathbb{R}}
\newcommand{\Sset}{\mathbb{S}}

\newcommand{\eV}{\vect{e}}              
\newcommand{\jV}{{\vect{j}}}		
\newcommand{\pV}{{\vect{p}}}            
\newcommand{\uV}{{\vect{u}}}            
\newcommand{\vV}{{\vect{v}}}            
\newcommand{\wV}{{\vect{w}}}            
\newcommand{\xV}{{\vect{x}}}            

\newcommand{\omV}{{\vect{\omega}}}      

\newcommand{\eVN}{{\vect{E}}}		
\newcommand{\uVN}{{\vect{U}}}     	
\newcommand{\vVN}{{\vect{V}}}     	
\newcommand{\wVN}{{\vect{W}}}           
\newcommand{\wVk}{{\vect{w}_k}}          
\newcommand{\wVl}{{\vect{w}_l}}          

\newcommand{\id}{\mathbf{I}}
\newcommand{\ze}{\mathbf{0}}

\newcommand{\dd}{\mathrm{d}}

\newcommand{\pdtau}{\partial_\tau}

\newcommand{\pdwk}{\partial_{\wV_k}}
\newcommand{\pdwl}{\partial_{\wV_l}}

\newcommand{\abs}[1]{\left| #1 \right|}

\renewcommand{\leq}{\leqslant}
\renewcommand{\geq}{\geqslant}

\newcommand{\vareps}{\varepsilon}

\newcommand{\smallr}{\sqrt{\scriptstyle{{2N\vareps_0}}}}
\newcommand{\intgpart}{\big\lfloor\!\frac{{\scriptstyle s}\!
{\scriptscriptstyle -} \!{\scriptstyle r}}{2}\!\big\rfloor}

\begin{document}

\title{
The linear Fokker-Planck equation for the 
Ornstein-Uhlenbeck~process~as~an~(almost)~nonlinear
kinetic equation for an isolated N-particle system}
\author{
Michael Kiessling\footnote{Department of Mathematics, 
Rutgers University, Piscataway NJ 08854} 
\ and 
Carlo Lancellotti\footnote{Department of Mathematics, 
City University of New York-CSI, Staten Island NY 10314} 
}

\date{Nov.2, 2005}
\maketitle


\begin{abstract}
\noindent
 It is long known that the Fokker-Planck equation with prescribed
constant coefficients of diffusion and linear friction 
describes the ensemble average of the stochastic evolutions in velocity space
of a Brownian test particle immersed in a heat bath of fixed temperature.
 Apparently, it is not so well known that the same partial differential 
equation, but now with constant coefficients which are functionals of 
the solution itself rather than being prescribed, 
describes the kinetic evolution (in the $N\to\infty$ limit) of an 
\textit{isolated} $N$-particle system with certain stochastic interactions.
 Here we discuss in detail this recently discovered interpretation.
\end{abstract}



\baselineskip=20pt

\section{Introduction}

As is well known,
\cite{UhlOrn}, \cite{Chandra}, \cite{Bal},
the ensemble average of the stochastic evolutions in {velocity space} 
of a Brownian test particle\footnote{For the beginnings of the theory of
	Brownian motion, see the collection of Einstein's papers
	with commentary \cite{Einstein}.}
of unit mass, 
immersed in a drifting uniform heat bath of fixed temperature $T$ and
constant drift velocity $\uV$, is governed by the Fokker--Planck equation 
with prescribed constant coefficients of diffusion and (linear) friction, 
\begin{equation}
\partial_t f (\vV;t)
=
\partial_{\vV}\cdot\Big({T}\partial_{\vV}f(\vV;t)
+ 
\big(\vV - \uV\big) f(\vV;t)\Big).
\label{FPbrown}
\end{equation}
 Here, $f(\, .\, ;t):\Rset^3\to\Rset_+$ is the ensemble's probability density 
function on velocity space at time $t\in \Rset_+$, and an overall constant
has been absorbed in the time scale. 
 Of course, we could also shift $\vV$ to obtain $\uV=\ze$, then 
rescale $\vV$, $t$, and $f$ to obtain $T = 1$; 
however, for pedagogical purposes we refrain from doing so. 
 The solution $f (\vV;t)$ of \refeq{FPbrown} is  given by
$f (\vV;t) = \int_{\Rset^3}G_t(\wV,\vV|\uV;T)f_0(\wV)\dd^3\wV$, 
where $f_0(\vV)\equiv f (\vV;0)$ and
\begin{equation}
G_t(\wV,\vV|\uV;T)
=
\left({2\pi T(1-e^{-2t})}\right)^{-\frac{3}{2}} 
\exp\left( -\frac{1}{2 T}\frac{|\vV-\uV -\wV e^{-t}|^2}{1-e^{-2t}} \right)
\label{OUkernel}
\end{equation}
is the Green function for \refeq{FPbrown}, see \cite{UhlOrn}, \cite{Chandra}.
 In its standard form, i.e. with 
$ T = 1$ and $\uV=\ze$, \refeq{OUkernel} is known as the (Mehler)
kernel of the adjoint Ornstein-Uhlenbeck semigroup 
(a.k.a. Fokker--Planck semigroup).

 Over the years, the Ornstein-Uhlenbeck semigroup and its adjoint
have come to play an important 
role in several branches of probability theory \cite{Hsu} related,
in some form, to Brownian motions.
 The fact that the explicitly known kernel \refeq{OUkernel} 
of the Fokker--Planck semigroup readily lends itself to analytical 
estimates has led to useful applications also outside the realm
of probability theory. 
  In particular, in recent years the Fokker--Planck semigroup has found 
applications in kinetic theory, the subfield of transport theory 
which is concerned with the approach to equilibrium and the response to 
driving external forces of individual continuum systems not in 
local thermal equilibrium; see, for instance, the review \cite{Vil}.  

 However, the linear Fokker--Planck equation itself, \refeq{FPbrown},
usually is not thought of as a {kinetic} equation for 
the particle density function on velocity space of an 
\textit{individual, isolated} space-homogeneous system of particles in 
some compact domain, which perform a microscopic autonomous dynamics 
that may be deterministic or stochastic but should satisfy the usual 
conservation laws of mass (particle number), energy and,
depending on the shape of the domain in physical space 
and its boundary conditions, also momentum and angular momentum.
 Evidently the very meaning of $f$ and the parameters $\uV$ and $T$ in 
\refeq{FPbrown} voids this interpretation. 
 Yet, with a re-interpretation of $f$,  $\uV$ and $T$ it \emph{is}
possible to assign to \refeq{FPbrown} a kinetic meaning. 

 Incidentally, the first result showing that at least a partial 
re-interpretation of \refeq{FPbrown} in this direction is
possible can be found in a paper by Villani \cite{Vil98} who, in his study 
of the space-homogeneous Landau equation for the weak deflection (i.e. Landau)
limit of a gas of particles with Maxwellian molecular interactions, discovered
that for isotropic velocity distribution functions $f$ (and only for these) 
the Landau equation is identical to \refeq{FPbrown}, with parameters 
$\uV=\vect{0}$ and $T$ matched to guarantee energy conservation.
 For general non-isotropic data the Landau equation for Maxwell
molecules is identical to a more complicated equation than \refeq{FPbrown}.
 
 To pave the ground for a complete re-interpretation of \refeq{FPbrown},
which requires re-assigning the meaning of $f$, $\uV$ and $T$, we first 
note that by the linearity of \refeq{FPbrown} we can scale $f$ to any
positive normalization we want. 
 We now introduce the following functionals of~$f$, 

\noindent
the ``mass of $f$''
\begin{equation}
m(f) = \int_{\Rset^3} f(\vV;t)\dd^3\vV\,,
\label{mOFf}
\end{equation}
the ``momentum of $f$''
\begin{equation}
\pV(f) = \int_{\Rset^3} \vV f(\vV;t)\dd^3\vV\,,
\label{pOFf}
\end{equation}
and the ``energy of $f$''
\begin{equation}
e(f) = \int_{\Rset^3} \frac{1}{2}|\vV|^2 f(\vV;t)\dd^3\vV\,.
\label{eOFf}
\end{equation}
 The ``angular momentum of $f$'' for a space-homogeneous $f(\vV;t)$
is simply $\jV(f) = \xV_{\mathrm{CM}}\times\pV(f)$, with $\xV_{\mathrm{CM}}$ 
the center of mass of the system, but this does not add any further insight 
and hence will not be considered explicitly.
 The functionals  \refeq{mOFf}, \refeq{pOFf}, and \refeq{eOFf}
inherit some time dependence from
the solution $f(\,.\,;t)$ of \refeq{FPbrown}, but to find 
this dependence explicitly it is not necessary to solve for $f$ first.
 Indeed, it is an elementary exercise in integration by parts 
to extract from \refeq{FPbrown} the following linear 
evolution equations with constant coefficients for $m$, $\pV$, and $e$,
\begin{equation}
\dot{m} = 0\,, 
\label{mDOT}
\end{equation}
\begin{equation}
\dot{\pV} = m\uV - \pV\,,
\label{pDOT}
\end{equation}
\begin{equation}
\dot{e} = 3 T - 2e + \uV\cdot\pV\,,
\label{eDOT}
\end{equation}
\newpage
\noindent
which, beside the conservation of mass, i.e. $m(f) = m(f_0)$, 
describe the exponentially fast convergence to a stationary state 
$\pV(f) \leadsto m(f_0)\uV$ and 
$e(f)\leadsto \frac{3}{2} T+ \frac{1}{2}m(f_0)|\uV|^2$.
  While all this is of course quite trivial and well known,
the relevant fact to realize here is that whenever the 
energy and the momentum of the initial $f_0$ equal these asymptotically
stationary values, viz. if  $\pV(f_0) = m(f_0)\uV$ and 
$e(f_0) = \frac{3}{2} T+ \frac{1}{2}m(f_0)|\uV|^2$, then beside 
mass $m$, also energy $e$ and momentum $\pV$ will be conserved.
 Conservation of mass, energy, and momentum for such a large subset
of initial data $f_0$ does not yet mean that we may already think 
of the linear 
equation \refeq{FPbrown} as a kinetic equation, which should 
conserve mass, energy, and (depending on the shape of the domain 
in physical space and its boundary conditions) also momentum
for \emph{all} 
initial data, no matter what their mass, energy and momentum are;
moreover, a genuine kinetic equation for particles with (pair or higher order) 
interactions must express the time derivative of $f$ in terms of an at 
least\footnote{The Boltzmann, the Landau, and the Vlasov kinetic 
	equations have bilinear ``interaction operators,'' 
	the Balescu--Lenard--Guernsey equation has a higher order 
	nonlinearity which reduces to the bilinear format in the 
	long wavelength regime.}
bilinear operator in $f$. 
 However, with the help of \refeq{mOFf}, \refeq{pOFf} and \refeq{eOFf} 
we now replace $T$ and $\uV$ in \refeq{FPbrown} to obtain just such a
kinetic equation.

 Indeed, consider the {\apriori} nonlinear Fokker--Planck equation
\begin{equation}
\partial_t f (\vV;t)
=
\partial_{\vV}\cdot\Big(\frac{1}{3}\big(2e(f)m(f)- |\pV(f)|^2\big)
\partial_{\vV}f(\vV;t)
+
\big(m(f)\vV - \pV(f)\big) f(\vV;t)\Big),
\label{FPkin}
\end{equation}
where $f(\, .\, ;t):\Rset^3\to\Rset_+$ now is a particle density function 
on velocity space at time $t\in \Rset_+$.
 The right-hand side of \refeq{FPkin} is a sum of a bilinear and a trilinear
operator acting on $f$ which now guarantees conservation of mass, momentum,
and energy for \emph{all} initial data $f_0\geq 0$, as verified by repeating 
the easy exercise in elementary integrations by parts using 
\refeq{FPkin} to find 
$\dot{m} = 0$ as well as $\dot{\pV} = m\pV - \pV m = \ze$ and 
$\dot{e} = 2em - |\pV|^2 m - 2em + |\pV|^2m = 0$.
 Of course, \emph{after this fact} of mass, momentum, and energy conservations
the {\apriori}  nonlinear equation \refeq{FPkin} 
in effect becomes just a completely and explicitly solvable
linear\footnote{In this sense \refeq{FPkin} is
	``almost nonlinear,'' or ``essentially linear,''
        depending on one's viewpoint.}
Fokker--Planck equation \refeq{FPbrown}, only now with parameters
$\uV$ and $T$ which are not prescribed but 
determined through the initial data $f_0$, viz.
$\uV = \pV(f_0)/m(f_0)\equiv \uV_0$ and 
$\frac{3}{2} T = e(f_0) - |\pV(f_0)|^2/2m(f_0)\equiv \vareps_0$;
we also set $m(f_0)\equiv m_0$ and $e(f_0) = e_0$.
 Accordingly, \refeq{FPkin}
inherits from \refeq{FPbrown} the feature that,
as $t\to\infty$, its solutions $f$ converge pointwise 
exponentially fast to the Maxwellian equilibrium state
\begin{equation}
f_{\mathrm{M}}(\vV) = 
m_0 \left(\frac{3}{4\pi\vareps_0}\right)^{\frac{3}{2}} 
\exp\left( -\frac{3|\vV -\uV_0|^2}{4\vareps_0} \right),
\end{equation}
with monotonically increasing relative entropy
\begin{equation}
S(f|f_{\mathrm{M}}) 
= 
- \int_{\Rset^3} f(\vV;t)\ln \frac{f(\vV;t)}{f_{\mathrm{M}}(\vV)}\dd^3\vV
\end{equation}
which in fact approaches its maximum value $0$ exponentially fast.

  Since  (\ref{FPkin})  displays all the 
familiar features of a  kinetic equation (formal nonlinearity;
conservation laws of mass, momentum, energy; an $H$-Theorem; 
approach to equilibrium; Maxwellian equilibrium states), at this point 
we may legitimately contemplate \refeq{FPkin} as a kinetic equation of 
some spatially homogeneous, isolated system of $N$ interacting particles 
in a compact spatial domain compatible with momentum conservation (e.g. a 
rectangle with periodic boundary conditions).
 In the remainder of this paper we show explicitly how \refeq{FPkin} 
arises from the Kolmogorov equation\footnote{In the physics literature, 
	the Kolmogorov equation for an $N$-particle Markov process is 
	traditionally called ``master equation".}
for the adjoint evolution of an underlying $N$-particle Markov process
in the limit $N\to\infty$. 
 We use the strategy originally introduced by Kac \cite{Kac} in 1956 
in the context of his work on a caricature of the Boltzmann equation;
for important recent work on Kac's original program,
see \cite{CarLoss}.
  As Kac realized, the crucial property that needs
to be established in order to validate the $N\to\infty$ limit is 
what he called ``propagation of chaos," which loosely speaking 
means that if the particle velocities are uncorrelated at $t=0$, 
they remain uncorrelated at later times; 
this can be rigorously true only on the continuum scale
in the limit $N\to\infty$. 

  Interestingly enough, by adding some suitable lower order terms to the 
putatively simplest $N$-particle Markov process that leads to the
(kinetic) Fokker--Planck equation in the limit $N\to\infty$, the corresponding
Kolmogorov equation for an ensemble of such isolated $N$-particle systems 
can be simplified to be just the diffusion equation on the $3N-4$-dimensional 
manifold (a sphere) of constant energy and momentum.
  Since therefore both the finite-$N$ and the infinite-$N$ equations are 
exactly solvable, the kinetic limit $N\to\infty$ can be carried out explicitly
and studied in great detail. 
 For this reason we actually defer the discussion of the underlying 
$N$-particle process to Appendix Ab while in the main part of our paper
we analyse the diffusion equation on $\Sset^{3N-4}_{\sqrt{2N\vareps_0}}$ and derive 
from it the kinetic Fokker--Planck equation on $\Rset^3$.

 Technically, we apply the Laplace--Beltrami operator to a probability
density on $\Sset^{3N-4}_{\sqrt{2N\vareps_0}}$ and then integrate out $N-n$ 
velocities over their constrained domain of accessibility. 
 Taking next the limit $N\to\infty$ yields a Fokker--Planck 
operator acting on the $n$-th marginal density on $\Rset^{3n}$.
 Thus we obtain a linear Fokker--Planck hierarchy of equations indexed by $n$.
 Using the Hewitt--Savage decomposition theorem, the hierarchy is seen to
be generated by the single, {\apriori}
 nonlinear kinetic Fokker--Planck equation 
\refeq{FPkin} which in view of the conservation laws is equivalent to the 
essentially linear Fokker--Planck equation \refeq{FPbrown}
 with constant parameters which are determined by the initial data.
 
  Experts in probability theory may have noticed a similarity between the first
part of our program and what has been called the ``Poincar\'e limit'' 
\cite{Bak}; in fact, our approach is ``dual'' to Bakry's approach. 
  More specifically, Bakry \cite{Bak} has shown that the action of the 
Laplace--Beltrami operator for $\Sset^N_{\sqrt{N}}\hookrightarrow \Rset^{N+1}$ 
on a probability density function over a ``radial'' coordinate 
axis of $\Sset^N_{\sqrt{N}}$ becomes identical, in the limit $N\to\infty$,
to the action of the Ornstein--Uhlenbeck operator on the same density viewed as
a function over $\Rset$.  Obviously, whenever the ``radial" function is
obtained by taking the marginal of a probability density over
$\Sset^N_{\sqrt{N}}$, i.e. by integrating out the $N-1$ Cartesian
coordinates of the embedding space which are perpendicular to a
fixed ``radial'' direction, the Ornstein--Uhlenbeck operator acts on the 
limiting marginal density as $N\to\infty$.
 This relationship between the operators is reflected at the spectral level 
by the convergence of the whole structure of orthogonal eigenfunctions
of the Laplacian on $\Sset^N_{\sqrt{N}}$ (hyper-spherical harmonics) to the 
orthogonal eigenfunctions of the Ornstein--Uhlenbeck operator on $\Rset$ 
(Hermite polynomials multiplied by the square root of their Gaussian
weight function); one of the earliest works is \cite{Mehler}, while
more recent works on the Poincar\'e limit, containing 
interesting connections with the theory of Markov semigroups, 
are \cite{Bak} and \cite{BakMaz}.
 Our procedure is ``dual'' to Bakry's approach in the sense that we integrate
out subsets of the Cartesian variables of the embedding space \emph{after} 
having applied the Laplace--Beltrami operator to a probability density on the
high-dimensional sphere, thereby obtaining the \emph{adjoint} 
Ornstein--Uhlenbeck operator acting on the respective marginals; 
in addition, while Bakry considers only mass and energy conservation, we
consider conservation of mass, energy, and momentum.

 Incidentally, our work is not inspired by Bakry's works on the 
Poincar\'e limit, nor by Villani's discovery about the isotropic
evolution of the space-homogeneous Landau equation, about both of
which we learned only after our own findings. 
 Rather, our study of the diffusion equations on the $3N-C$-dimensional spheres
of constant energy ($C=1$), respectively energy and momentum ($C=4$), which
began in \cite{KieLan04}, was originally conceived of as a 
\emph{technically simpler primer} for our investigation (also in 
\cite{KieLan04}) of the Balescu--Prigogine master equation for 
Landau's kinetic equation. 
 And while the present paper is also a technical continuation of 
\cite{KieLan04}, in the sense that here we supply various calculations 
that we had announced in \cite{KieLan04}, the main purpose of the present 
paper is to amplify the conceptual spin-off of our technical investigations, 
the \emph{new physical interpretation} of one of the simplest and best known 
linear transport equations as an (almost nonlinear) kinetic equation.
 As should be clear from our discussion in this introduction, 
this kinetic theory interpretation of the prototype Fokker--Planck 
equation may have been suspected by others long ago, yet we have not 
been able to find the whole story in the literature. 

 In what follows, for the sake of simplicity we set $m_0 =1$, 
and accordingly\footnote{Setting $m_0=1$ means we should now 
	speak of the energy per particle $e_0$, the
	thermal energy per particle $\vareps_0$,
	and the momentum per particle $\pV_0 (=\uV_0)$.}
obtain $\pV(f_0) \equiv \uV_0$ and 
$e(f_0) - |\pV(f_0)|^2/2 = e_0 - |\uV_0|^2/2 \equiv \vareps_0$.
 With these simplifications \refeq{FPkin} now becomes
\begin{equation}
\partial_t f (\vV;t)
=
\partial_{\vV}\cdot\Big(\frac{2}{3}\vareps_0
\partial_{\vV}f(\vV;t)
+
\big(\vV - \uV_0\big) f(\vV;t)\Big).
\label{FPkinSIMPLE}
\end{equation}
 While \refeq{FPkinSIMPLE} is essentially a linear PDE, 
it should just be kept in mind that $\vareps_0$ and $\uV_0$
are functionals of $f$ which are determined by the initial data 
$f_0$ and not chosen 
independently.\footnote{The identification of \refeq{FPkin} with
	\refeq{FPkinSIMPLE} is valid only for isolated systems
	that can freely translate. If a driving external force
	field $\mathbf{F}$ is applied, then $e(f)$ and $\pV(f)$ 
	are no longer constant and \refeq{FPkin} -- with the addition
	of the forcing term $-\mathbf{F}\cdot\partial_\vV f$ to its r.h.s.
	-- is the relevant equation.}
 We next shall derive \refeq{FPkinSIMPLE} from the diffusion equation equation 
on $\Sset^{3N-4}_{\sqrt{2N\vareps_0}}$ in the spirit of Kac's program. 


\section{The Finite-$N$ Ensembles}


Consider an infinite ensemble of i.i.d. random vectors
$\{\vVN_\alpha\}_{\alpha =1}^\infty$ where each
$\vVN =(\vV_1,...,\vV_N)\in \Rset^{3N}$ represents a possible
micro-state of an individual system of $N$ particles with velocities
$\vV_i=(v_{i1},v_{i2},v_{i3})\in\Rset^3$ and particle positions assumed
to be uniformly distributed over a periodic box; hence, particle positions
will not be considered explicitly.
 Each $\vVN$ takes values in the $3N-4$-dimensional manifold of constant
energy $e_0$ and momentum $\uV_0$,
\begin{equation}
\Mset^{3N-4}_{\uV_0,e_0}
=
\Big\{\vVN\; :\;\sum_{k=1}^N \vV_k=N\mathbf{u}_0, \;
\sum_{k=1}^N\frac{1}{2} \abs{\vV_k}^2=Ne_0,
\; e_0 > \frac{1}{2}|\uV_0|^2 \Big\}.
\end{equation}
 The manifold $\Mset^{3N-4}_{\uV_0,e_0}$ is identical to a
$3N-4$-dimensional sphere of radius $\sqrt{2N\vareps_0}$
(where $\vareps_0$ appears above \refeq{FPkinSIMPLE}),
centered at
$\uVN = (\uV_0,...,\uV_0)$ and embedded in the $3(N-1)$-dimensional
affine linear subspace of $\Rset^{3N}$ given by $\uVN + \Lset^{3N-3}$, where
$\Lset^{3N-3} \equiv \Rset^{3N}\cap\big\{\vVN\in\Rset^{3N}:
\sum_{k=1}^N \vV_k= \mathbf{0}\big\}$
is the space of velocities in any center-of-mass frame.
 The ensemble at time $\tau$ is characterized by a probability density
$F^{(N)}(\vVN;\tau)$ on $\Mset^{3N-4}_{\uV_0,e_0}$, the evolution of which
is determined by the diffusion equation
\begin{equation}
\pdtau F^{(N)}(\vVN;\tau)
=
\Delta_{\Mset^{3N-4}_{\uV_0,e_0}}\,F^{(N)}(\vVN;\tau),
\label{heat}
\end{equation}
where $\Delta_{\Mset^{3N-4}_{\uV_0,e_0}}$ is the
Laplace--Beltrami operator on $\Mset^{3N-4}_{\uV_0,e_0}$.
 Since all particles are of the same kind, we consider only solutions to
 (\ref{heat}) which are invariant under the symmetric
group $S_N$ applied to the $N$ components in $\Rset^3$ of $\vVN$.
 Clearly, permutation symmetry is preserved by the evolution.\footnote{In
        what follows, for the sake of notational simplicity we will not
        enforce this symmetry explicitly, but the reader should be aware
        that (for instance) all the eigenfunctions that appear below in
        the solution for $F^{(N)}$ can be easily symmetrized.}
We will show that the diffusion equation (\ref{heat}), here viewed as
a master equation,  leads precisely to the essentially linear
Fokker--Planck equation \refeq{FPkinSIMPLE}
in the sense of Kac's program:
(a) the Fokker--Planck equation \refeq{FPkinSIMPLE}
arises as the $N\to\infty$ limit of the equation for the
first marginal of $F^{(N)}(\vVN;\tau)$ derived from
 (\ref{heat}), and (b) propagation of chaos holds.
 In this section we prepare the ground by discussing the finite-$N$
equation \refeq{heat}.
 The limit $N\to\infty$ is carried out in the next section, while
propagation of chaos is discussed in the final section.

 For the sake of completeness, we begin by listing some general
facts about the diffusion equation.
 We note that the Laplacian $\Delta_{\Mset^{3N-4}_{\uV_0,e_0}}$
is a positive semi-definite, essentially self-adjoint operator
on the dense domain
$\Csp^\infty (\Mset^{3N-4}_{\uV_0,e_0})
        \subset\Lsp^2 (\Mset^{3N-4}_{\uV_0,e_0})$,
thus it has a unique self-adjoint extension with domain
$\Hsp^2(\Mset^{3N-4}_{\uV_0,e_0})$.
 Its self-adjoint extension is the generator of a non-expansive semigroup
on $\Lsp^2 (\Mset^{3N-4}_{\uV_0,e_0})$ which is strictly contracting on
the $\Lsp^2$ orthogonal complement of the constant functions.
 Thus, we may ask that the initial condition
$\lim_{t\downarrow 0}F^{(N)}(\,.\,;\tau)
= F_0^{(N)}(\,.\,)\in\Lsp^2 (\Mset^{3N-4}_{\uV_0,e_0})$
(which implies $F_0^{(N)}\in \Lsp^1 (\Mset^{3N-4}_{\uV_0,e_0})$).
 Yet, as is well-known, the diffusion semigroup is so strongly regularizing
that we may even take
$F_0^{(N)}(\,.\,)\in\Msp_{+,1} (\Mset^{3N-4}_{\uV_0,e_0})$,
a probability measure, and obtain
$F^{(N)}(\,.\,;\tau) \in \Csp^\infty(\Mset^{3N-4}_{\uV_0,e_0})$
for all $\tau>0$.

 In fact, the solutions of  \refeq{heat} can be computed quite
explicitly in terms of an eigenfunction expansion.
 Since via translation by $\uVN$ (choosing a center-of-mass frame)
and scaling by $\sqrt{2N\vareps_0}$ (choosing a convenient unit of energy)
the manifold $\Mset^{3N-4}_{\uV_0,e_0}$ can be identified with the
unit sphere centered at the origin of the linear subspace
$\Lset^{3N-3}\subset\Rset^{3N}$, the complete spectrum of
$\Delta_{\Mset^{3N-4}_{\uV_0,e_0}}$ and an orthogonal basis
of eigenfunctions can be obtained from the well-known eigenvalues and
eigenfunctions for the Laplacian on the unit sphere
$\Sset^{3N-4}\hookrightarrow\Rset^{3N-3}$.
Of course, in our case the
embedding is $\Sset^{3N-4}\hookrightarrow\Lset^{3N-3}$ with
$\Lset^{3N-3}$ isomorphic by a rotation to standard $\Rset^{3N-3}$.
 Thus we start from $\Mset^{3N-4}_{\uV_0,e_0}$ and
we first carry out a rotation in $\Rset^{3N}$ that transforms $\vVN$ to
$\wVN=\mathcal{U}\vVN$ in such a way that $\Lset^{3N-3}$ is mapped to the
$3N-3$-dimensional linear subspace $\big\{\wVN\ :\ \wV_N=\ze\big\}$.
 Obviously, $\mathcal{U}^T$ is the linear transformation that diagonalizes
the projection operator onto $\Lset^{3N-3}$.
 A complete orthonormal set of eigenvectors
for such a projection is readily calculated and leads to
\begin{eqnarray}
\wV_1
&=&
\sqrt{\frac{N-1}{N}}\left[\vV_1-\frac{1}{N-1}\sum_{i=2}^N\vV_i\right]
\nonumber\\
&\vdots&\nonumber\\
\wV_n&=&\sqrt{\frac{N-n}{N-n+1}}\left[\vV_n-\frac{1}{N-n}\sum_{i=n+1}^N\vV_i
\right]\nonumber\\
&\vdots&\nonumber\\
\wV_{N-1}\!\!\!\!\!\!&=&\frac{1}{\sqrt{2}}[\vV_{N-1}-\vV_N]\nonumber\\
\wV_N&=&\frac{1}{\sqrt{N}}\sum_{i=1}^N\vV_i
\label{rotation}
\end{eqnarray}
 It is easily checked that the matrix associated with this transformation
is indeed orthogonal, and that $\wV_N$ vanishes whenever $\vVN\in\Lset^{3N-3}$.
 More generally, the
affine subspace $\uVN+\Lset^{3N-3}$ is mapped to the linear manifold
$\big\{\wVN\ :\ \wV_N=\sqrt{N}\uV_0\big\}$
and $\Mset^{3N-4}_{\uV_0,e_0}$ is mapped to
\begin{equation}
\bigg\{\wVN\; :\; \wV_N=\sqrt{N}\uV_0,\quad
\sum_{i=1}^{N-1}\abs{\wV_i}^2=2Ne_0-N\abs{\uV_0}^2=2N\vareps_0
\bigg\}
\end{equation}
which implies that the truncated vector $(\wV_1,\dots,\wV_{N-1})$ belongs
to the sphere $\Sset_{\sqrt{2N\vareps_0}}^{3N-4}\hookrightarrow\Rset^{3N-3}$
(in $\wV_k$-coordinates).
 Thus, the transform $\mathcal{U}$ allows one to analyse
the $N$-particle system with energy and momentum conservation
(``periodic box" setup) in terms of an $(N-1)$-particle system
with only energy conservation (a ``container with reflecting walls"
setup).\footnote{The gas in such a container was discussed in our
        earlier work \cite{KieLan04}, but without detailed calculations.
        Our calculations with the $\wV$ variables here now supply the
        relevant details.}
 For future reference, we also observe that for $n$ fixed and $N\to\infty$
the effect of $\mathcal{U}$ reduces to a translation of each of the $n$
velocities by $\uV_0$, in the following sense.
 Consider a consistent hierarchy of vectors of increasing size $N$, in
which lower-$N$ vectors can be obtained from the higher-N ones by truncation
(i.e. projection).
 Suppose that the vectors belong to $\uVN+\Lset^{3N-3}$ for all $N$, apply
the transformation in  (\ref{rotation}) and look at the $n$-th component.
 Since $\sum_{i=n+1}^N\vV_i=N\uV_0-\sum_{i=1}^n\vV_i$,
where $\sum_{i=1}^n\vV_i$ is independent of $N$, we find
\begin{equation}
\lim_{N\to\infty}\wV_n=\vV_n-\uV_0.
\label{wntovn}
\end{equation}

 We now recall that the Laplacian is invariant under Euclidean
transformations.
 Thus, under our orthogonal transformation $\mathcal{U}$,
the Laplacian $\Delta_{\Mset^{3N-4}_{\uV_0,e_0}}$ becomes the Laplacian on
$\Sset_{\sqrt{2N\vareps_0}}^{3N-4}$ in $\Rset^{3N-3}$, the space of
truncated vectors $(\wV_1,\dots,\wV_{N-1})$ (which will also be denoted
by $\wVN$, at the price of abusing the notation).
  Since $\Delta_{\Sset_{\sqrt{2N\vareps_0}}^{3N-4}}=
\frac{1}{2N\vareps_0}\Delta_{\Sset^{3N-4}}$, and
the Laplacian on the unit sphere $\Sset^{3N-4}$ has spectrum
$j(j + 3N -5)$, $j=0,1,\dots$,
the spectrum of $\Delta_{\Mset^{3N-4}_{\uV_0,e_0}}$ is
\begin{equation}
\lambda^{(j)}_{\Mset^{3N-4}_{\uV_0,e_0}}=
\frac{j(j + 3N -5)}{2N\vareps_0},\qquad j=0,1,\dots\ .
\medskip
\label{eigenvaluesSN}
\end{equation}
  The eigenspace on $\Sset^{3N-4}$ for the $j$-th eigenvalue has dimension
\begin{equation}
\mathcal{N}(j,3N-3)=\frac{(3N-5+2j)(3N-6+j)!}{j!(3N-5)!}
\end{equation}
and is spanned by an orthogonal basis of
hyper-spherical harmonics\footnote{The hyper-spherical harmonics
        on $\Sset^n$ are restrictions to $\Sset^n\subset \Rset^{n+1}$ of
        homogeneous harmonic polynomials in $\Rset^{n+1}$.
        For $j>0$ the restriction has to be non-constant, since
        $\widetilde{Y}_{0,1}\equiv\ const.$.}
on $\Sset^{3N-4}\subset \Rset^{3N-3}$ of order $j$, here
denoted $\widetilde{Y}_{j,\ell}(\omV;3N-3)$,
with $\ell\in\Dset_j=\{1,\dots,\mathcal{N}(j,3N-3)\}$
and with $\omV\in\Sset^{3N-4}$.
 The indexing of our $\widetilde{Y}_{j,\ell}(\omV;3N-3)$ follows the
convention of \cite{Mul} for his $Y_{j,\ell}$ and differs from what
might have been anticipated from the familiar convention for spherical
harmonics on $\Sset^2$.
Our reason for using tildes atop the function symbols is to remind
the reader that we will use a  normalization of the
$\widetilde{Y}_{j,\ell}(\omV;3N-3)$
which conveniently suits our purposes and does not seem to agree with any
of the existing conventions, such as in \cite{Mul} or for the spherical
harmonics on $\Sset^2$.
 Our convention is motivated by the analysis of the large
$N$ behavior of the eigenfunctions, carried out in Appendix B.

 Hence, the eigenspace of $\Delta_{\Mset^{3N-4}_{\uV_0,e_0}}$
associated with the $j$-th eigenvalue in  (\ref{eigenvaluesSN})
is spanned by the eigenfunctions
$\widetilde{Y}_{j,\ell}\left({\wVN}/{\sqrt{2N\vareps_0}};3N-3\right)$,
$\ell\in\Dset_j$,
where $\wVN$ is given by  (\ref{rotation}) for $n=1,\dots,N-1$.
 To shorten the notation we introduce
\begin{equation}
G_{j,\ell}^{(N)}(\vVN)
\equiv \abs{\Mset^{3N-4}_{\uV_0,e_0}}^{-1}
\widetilde{Y}_{j,\ell}\left({\wVN}/{\sqrt{2N\vareps_0}}\,;3N-3\right);
\label{eigenfunctions}
\end{equation}
here, the factor $\abs{\Mset^{3N-4}_{\uV_0,e_0}}^{-1}$ is introduced
for later convenience.

 In terms of the eigenfunctions $G_{j,\ell}^{(N)}(\vVN)$, the solution
to equation \refeq{heat} is simply given by the generalized Fourier series
\begin{equation}
F^{(N)}(\vVN;\tau)
=
\abs{\Mset^{3N-4}_{\uV_0,e_0}}^{-1}
+
\sum_{j\in\Nset} \sum_{\ell\in\Dset_j}
F_{j,\ell}^{(N)}
G_{j,\ell}^{(N)}(\vVN)\,
e^{- \textstyle{\frac{j(j +3N -5)}{2N\vareps_0}}\tau}
\label{FNevolution}
\end{equation}
with Fourier coefficients $F_{j,\ell}^{(N)}$ given by
\begin{equation}
F_{j,\ell}^{(N)}
=
\frac{\langle F^{(N)}_0|G_{j,\ell}^{(N)}\rangle}
     {\langle G^{(N)}_{j,\ell}|G_{j,\ell}^{(N)}\rangle}
\label{FourierCOEFF}
\end{equation}
where $\langle\,.\,|\,.\,\rangle$ denotes the inner product in
$\Lsp^2(\Mset^{3N-4}_{\uV_0,e_0})$.
 Notice, though, that the numerator 
$\langle F^{(N)}_0|G_{j,\ell}^{(N)}\rangle$
can be extended to mean the canonical pairing of the
$G_{j,\ell}^{(N)}$s with an element of their dual space,
which allows us to take $F^{(N)}_0$ to be a measure.
 In particular, we may take $F^{(N)}_0$ to be the Dirac measure concentrated
at any particular point of $\Mset^{3N-4}_{\uV_0,e_0}$.
 The formula \refeq{FNevolution} then describes the fundamental solution
of the diffusion equation \refeq{heat}.
 In any event, whatever $F^{(N)}_0$, \refeq{FNevolution} makes it evident
that when $\tau\to\infty$ the ensemble probability density function
on $\Mset^{3N-4}_{\uV_0,e_0}$ decays exponentially fast to the
uniform probability density $\abs{\Mset^{3N-4}_{\uV_0,e_0}}^{-1}=
        \abs{\Sset_{\sqrt{2N\vareps_0}}^{3N-4}}^{-1}=
F_{0,1}^{(N)} G_{0,1}^{(N)}(\vVN)$, which is the constant eigenfunction
corresponding to the smallest non-degenerate eigenvalue $0$ of the Laplacian.


\section{Evolution of the Marginals}


 To study the limit $N\to\infty$ for the time-evolution of the
ensemble measure, we need to consider the hierarchy of $n$-velocity 
marginal distributions 
\begin{equation}
F^{(n|N)}(\vV_1,\dots,\vV_n;\tau)
\equiv
\int_{\Omega^{3(N-n)-4}_{\mathbf{u}_0,e_0}} 
F^{(N)}(\vVN;\tau)\, d\vV_{n+1}\dots d\vV_N
\end{equation}
where $\Omega^{3(N-n)-4}_{\mathbf{u}_0,e_0}$ is
given by all the $(\vV_{n+1},\dots ,\vV_N)$ such that
\begin{equation}
\sum_{i=n+1}^N\!\! \vV_k=N\uV_0-\sum_{i=1}^n\vV_k,
\quad \sum_{i=n+1}^N \abs{\vV_k}^2=2Ne_0-
\sum_{i=1}^n \abs{\vV_k}^2  
\end{equation}
and $F^{(n|N)}$ has domain $\{(\vV_1,\dots,\vV_n):
\sum_{k=1}^n |\vV_k-\uV_0|^2\leq {4(N-n)\vareps_0}\}\subset\Rset^{3n}$.
 The evolution equation for the $n$-th marginal 
$F^{(n|N)}(\vV_1,\dots,\vV_n;\tau)$ is obtained by integrating 
\refeq{heat} over $(\vV_{n+1},\dots,\vV_N)\in \Rset^{3N-3n}$, 
using the representation of the Laplace--Beltrami operator given in 
(\ref{heat1}) of Appendix Aa.
  Then, a straightforward calculation 
(previously presented in \cite{KieLan04}) shows that $F^{(n|N)}$ satisfies
\begin{eqnarray}
\pdtau F^{(n|N)}\!\!&=&\!\!
\sum_{i=1}^{n}
\frac{\partial}{\partial\vV_i}\cdot\frac{\partial F^{(n|N)}}{\partial\vV_i}-
\frac{1}{N}\sum_{k=1}^3\sum_{i,j=1}^{n}
\frac{\partial^2 F^{(n|N)}}{\partial v_{ik}\partial v_{jk}}
\nonumber\\
&&
-\frac{1}{2N\vareps_0}
\sum_{i,j=1}^{n}\frac{\partial}{\partial\vV_i}\cdot
\left((\vV_i-\uV_0)\,(\vV_j-\uV_0)\cdot
\frac{\partial F^{(n|N)}}{\partial\vV_j}\right)
\nonumber\\
&&
+\frac{3(N-n)}{2\vareps_0N}\sum_{i=1}^{n}
\frac{\partial}{\partial\vV_i}\cdot\Big((\vV_i-\uV_0) F^{(n|N)}\Big).
\label{nDIFFhierarchyEQ}
\end{eqnarray}

 Clearly, to obtain the solutions of these equations it is advisable
to integrate the series solution for $F^{(N)}(\vVN;\tau)$, 
 (\ref{FNevolution}).
 For this purpose, it will be convenient to calculate the marginals 
in terms of the rotated variables $\wVN$.  
 Changing the integration variables\footnote{Note that 
	 (\ref{rotation}) defines a one-to-one linear map 
	with determinant $\sqrt{\frac{N}{N-n}}$ between
	$(\vV_{n+1},\dots\vV_N)$ and $(\wV_{n+1},\dots\wV_{N-1},\vect{z}_N)$,
	where $\vect{z}_N\equiv\wV_N-\frac{1}{\sqrt{N}}\sum_{i=1}^n\vV_i$.} 
gives
\begin{equation}
F^{(n|N)}(\vV_1,\dots,\vV_n;\tau)
=
{\textstyle{\sqrt{\frac{N}{N-n}}}}
\int F^{(N)}(\vVN;\tau)\,d\wV_{n+1}\dots d\wV_{N-1}
\label{marginW}
\end{equation}
where the integral is over 
$\Sset^{3(N-n)-4}_{\sqrt{2N\vareps_0-\sum_{i=1}^n\abs{\wV}_i^2}}$,
and we abused the notation $F^{(N)}(\vVN;\tau)$ by applying it to what is now 
regarded as a function of $(\vV_1,...,\vV_n,\wV_{n+1},...,\wV_{N-1})$.
 To obtain the series solution for $F^{(n|N)}(\vVN;\tau)$ we need to
express  (\ref{FNevolution}) in the variables 
$(\vV_1,...,\vV_n,\wV_{n+1},...,\wV_{N-1})$ and then integrate
term by term in the spirit of \refeq{marginW}.
 To accomplish this we need to choose explicitly a basis of 
spherical harmonics $\widetilde{Y}_{j,\ell}$ on $\Sset^{3N-4}$.  
 It is convenient to do this 
in an iterative fashion, by assuming that a basis is known for the
spherical harmonics with one independent variable less, 
here $\widetilde{Y}_{k,m}(\omV_{3N-5};3N-4)$ with
$\omV_{3N-5}\in\Sset^{3N-5}$. 
 Then, the desired basis is obtained \cite{Mul}
by taking all the elements in the given lower-dimensional basis and 
multiplying them by associated Legendre functions of the ``extra"
variable.  
 In our case the $(3N-3$)-th variable will be $w_{11}/\smallr$, the first 
component of $\wVN/\smallr$, and $\omV_{3N-5}$ will be 
a unit vector in the space of the remaining $3N-4$ components,
denoted by $(\wVN)_{3N-4}/\smallr$; thus, 
\begin{equation}
\widetilde{Y}_{j,\ell}\left(\frac{\wVN}{\smallr};3N-3\right)
=
\widetilde{Y}_{k,m}\left(\frac{(\wVN)_{3N-4}}{\smallr};3N-4\right)
\,
\widetilde{P}_j^k\left(\frac{w_{11}}{\smallr};3N-3\right)
\label{basis1}
\end{equation}
where $k=0,1,\dots, j$, $m=1,\dots,\mathcal{N}(k,3N-4)$ and each choice
of the pair $k,m$ is associated with a value of the degeneracy index 
$\ell$ for the basis $\widetilde{Y}_{j,\ell}$; moreover,
$\widetilde{P}_j^k$ is an associated Legendre 
function \cite{Mul}, suitably normalized (see Appendix B). 
 By repeating this process $3n$ times, we write out the eigenfunctions
in the form
\begin{eqnarray}
&&\hskip-.9truecm
\widetilde{Y}_{j,\ell}\left(\frac{\wVN}{\smallr};3N-3\right) 
=
\widetilde{Y}_{k_{3n},m}\left(\frac{(\wVN)_{3N-3n-3}}{\smallr};3N-3n-3\right)
\times\phantom{spaaace}
\\[0.4cm]
&\times&\!\!\!\!\!\!
\widetilde{P}_{k_{3n-1}}^{k_{3n}}\!\!
                        \left(\frac{w_{n3}}{\smallr};3N-3n-2\right)
\widetilde{P}_{k_1}^{k_2}\!\left(\frac{w_{12}}{\smallr};3N-4\right)
\cdots
\widetilde{P}_j^{k_1}\!\left(\frac{w_{11}}{\smallr};3N-3\right)
\nonumber
\label{basis2}
\end{eqnarray}
where $0\leq k_{3n}\leq \dots\leq k_1\leq j$ and 
$m=1,\dots,\mathcal{N}(k_{3n},3N-3n-3)$.
 Now let $g_{j,\ell}^{(n|N)}(\vV_1,\dots,\vV_n)$ denote the 
$n$-th ``marginal'' of $G_{j,\ell}^{(N)}(\vVN)$ 
(as for $F^{(N)}$ in  (\ref{marginW})), and
set $N^*\equiv N - n -1$. 
 We find
\begin{eqnarray}
g_{j,\ell}^{(n|N)}(\vV_1,\dots,\vV_n)
=
\abs{\Mset^{3N-4}_{\uV_0,e_0}}^{-1}
\int \widetilde{Y}_{k_{3n},m}\left(\frac{(\wVN)_{3N^*}}
				   {\smallr};3N^*\right)
d\wV_{n+1}\dots d\wV_{N-1}
\nonumber\\[0.4cm]
\times
{\textstyle{\sqrt{\frac{N}{N-n}}}}
\widetilde{P}_{k_{3n-1}}^{k_{3n}}\!\!
		\left(\frac{w_{n3}}{\smallr};3N\!-\!3n\!-\!2\right)\!\!\!
\cdots
\widetilde{P}_j^{k_1}\!\left(\frac{w_{11}}{\smallr};3N\!-\!3\right)
\end{eqnarray}
where the integral is over the same domain as in \refeq{marginW}.
  The integral of $\widetilde{Y}_{k_{3n},m}$ is non-zero if and only 
if $k_{3n}=0$ and $m=1$, and the integrals over $\widetilde{Y}_{0,1}$ 
are determined only up to the overall factor $\widetilde{Y}_{0,1}$, 
which we may choose to be unity without loss of generality.
  Accordingly, $g_{j,\ell}^{(n|N)}\equiv 0$ unless 
$\ell\in\widetilde\Dset_j\subset\Dset_j$, where 
$\widetilde\Dset_j$ contains the indices of the basis functions that 
``descend'' from the uniform harmonic in $\Rset^{3N-3n-3}$.
 For such $\ell$'s the integrated eigenfunctions then become 
\begin{eqnarray}
g_{j,\ell}^{(n|N)}(\vV_1,\dots,\vV_n)
=
\widetilde{P}_j^{k_1}\!\left(\frac{w_{11}}{\smallr};3N\!-\!3\right)
\cdots
\widetilde{P}_{k_{3n-1}}^{0}\!\!\left(\frac{w_{n3}}{\smallr},
3N\!-\!3n\!-\!2\right)
\nonumber\\[0.4cm]
\times
\sqrt{\frac{N}{N-n}}\,
\frac{\abs{\Sset^{3(N-n)-4}}}
     {\abs{\Sset^{3N-4}}}
\frac{1}{{\smallr}^{3n}}\,  
\Big(1-\frac{1}{\smallr}\sum_{i=1}^n|\wV_i|^2\Big)^{\frac{3(N-n)-4}{2}}.
\label{eigmarg2}
\end{eqnarray}
  The series for the $n$-th marginal $F^{(n|N)}(\,.\,;\tau)$ (the integrated 
\refeq{FNevolution}) is a series in the functions \refeq{eigmarg2}, viz.
\begin{equation}
F^{(n|N)}(\vV_1,\dots,\vV_n;\tau)
=
\sum_{j\in \Nset\cup\{0\}} \sum_{\ell\in\widetilde\Dset_j}
F_{j,\ell}^{(N)}
g_{j,\ell}^{(n|N)}(\vV_1,\dots,\vV_n)\,
e^{- \textstyle{\frac{j(j +3N -5)}{2N\vareps_0}}\tau}.
\label{FnNevolution}
\end{equation}


\section{The Limit $N\to\infty$}


We are now ready to take the infinitely many particles limit.
 First of all, we observe that the evolution equation for
the marginal velocity densities
$f^{(n)}(\vV_1,\dots,\vV_n;\tau)\equiv
\lim_{N\to\infty}F^{(n|N)}(\vV_1,\dots,\vV_n;\tau)$
which obtains in the formal limit $N\to\infty$ from \refeq{nDIFFhierarchyEQ}
is the essentially linear Fokker--Planck equation in $\Rset^{3n}$,
\begin{equation}
\pdtau f^{(n)}
=
\sum_{i=1}^{n}
\frac{\partial}{\partial\vV_i}\cdot\Big(\frac{\partial f^{(n)}}
{\partial\vV_i}+\frac{3}{2\vareps_0}(\vV_i-\uV_0)\,f^{(n)}\Big).
\label{nDIFFhierarchyEQlim}
\end{equation}
 We now show that the series expansion for the time-evolved
finite-$N$ marginals $F^{(n|N)}(\,.\,;\tau)$ converge under
natural conditions to solutions of these equations.

 Beginning with the spectrum of $\Delta_{\Mset^{3N-4}_{\uV_0,e_0}}$,
we note that the limit $N\to\infty$ yields
\begin{equation}
  \lim_{N\to\infty}
\Bigl\{
        \lambda_{\Mset^{3N-4}_{\uV_0,e_0}}^{(j)}
\Bigr\}_{j=0}^\infty
=
\Big\{\textstyle{
        \frac{3j}{2\vareps_0}}
\Big\}_{j=0}^\infty.
\label{LIMspectrumNEw}
\end{equation}
 Thus, the limit spectrum is discrete.
 In particular, there is a spectral
gap separating the origin from the rest of the spectrum.
 As a result, the time evolution of the limit $N\to\infty$
continues to approach a stationary state exponentially
fast when $\tau\to\infty$.

 Coming to the eigenfunctions, the expression
on the second line in  (\ref{eigmarg2}) contains the
$n$-velocity marginal distribution of the uniform density
$\abs{\Mset^{3N-4}_{\uV_0,e_0}}^{-1}$ (the $j=0$ case).
 As is well-known at least since the time of Boltzmann,
this distribution converges pointwise when $N\to\infty$
to the $n$-velocity drifting Maxwellian on $\Rset^{3n}$,
\begin{equation}
f_{\mathrm{M}}^{\otimes{n}}(\vV_1,...,\vV_n)
=
\left({\frac{3}{4\pi\vareps_0}}\right)^{\frac{3n}{2}}
\prod_{i=1}^n
\exp\left( -{\textstyle{\frac{3}{4\vareps_0}}}|\vV_i-\uV_0|^2 \right)
\label{nMaxwellian}
\end{equation}
(recall  (\ref{wntovn})).
 In terms of eigenfunctions this means that the ``projection'' onto
$\Rset^{3n}$ of the $j=0$ eigenfunction of the Laplace--Beltrami operator
on $\Sset^{3N-4}_{\sqrt{2N\vareps_0}}$ converges pointwise (in fact,
even uniformly) to the $j=0$ eigenfunction of the linear Fokker--Planck
operator in $\Rset^{3n}$, appearing in the r.h.s. of
 (\ref{nDIFFhierarchyEQlim}).
 The connection between the eigenfunctions generalizes to the cases
$j\neq 0$; cf. \cite{BakMaz} for the special case $\uV_0=\vect{0}$.
  The asymptotic behavior for $N\to\infty$ of
the associated Legendre functions in  (\ref{eigmarg2}),
which is discussed in Appendix B, together with  (\ref{wntovn}),
yields that
$g_{j,\ell}^{(n)}(\vV_1,\dots,\vV_n) \equiv
\lim_{N\to\infty}g_{j,\ell}^{(n|N)}(\vV_1,\dots,\vV_n)$
exists pointwise for all $(\vV_1,\dots,\vV_n)\in\Rset^{3n}$, with
\setlength{\arraycolsep}{0.5mm}
\begin{eqnarray}
\!\!\!\!\!\!
g_{j,\ell}^{(n)}(\vV_1,\dots,\vV_n)
&=&
\frac{(-1)^j}{2^{j/2}}
H_{j-k_1}\!\left({\textstyle{\sqrt{\frac{3}{4\vareps_0}}}}
(v_{11}\!-u_1)\!\right)
\cdots
H_{k_{3n-1}}\!\left({\textstyle{\sqrt{\frac{3}{4\vareps_0}}}}
(v_{n3}\!-u_3)
\!\right)
\label{eigmarg3}
\nonumber\\
&&\times
\left({\frac{3}{4\pi\vareps_0}}\right)^{\frac{3n}{2}}\,
\prod_{i=1}^n
\exp\left( -{\textstyle{\frac{3}{4\vareps_0}}}|\vV_i-\uV_0|^2 \right)
\nonumber\\
&\equiv&\frac{(-1)^j}{2^{j/2}}\!
\left({\textstyle{\frac{3}{4\pi\vareps_0}}}\right)^{\!\!\frac{3n}{2}}\!
\prod_{i=1}^n
e^{-\frac{3}{4\vareps_0}|\vV_i-\uV_0|^2}
\prod_{l=1}^3
H_{m_{i\cdot l}}\!\left({\textstyle{\sqrt{\frac{3}{4\vareps_0}}}}
(v_{il}\!-u_l)\!\right)
\label{nEIGlim}
\end{eqnarray}
for all $\ell\in\widetilde\Dset_j$,
where $H_m(x)$ is the Hermite polynomial of degree $m$ on $\Rset$, and
we defined $m_1=j-k_1,m_2=k_1-k_2,\dots, m_{3n}=k_{3n-1}$.  In terms of
the $m_i$'s, the index set $\widetilde\Dset_j$ counts all the choices of
integers
$0\leq m_1,\dots,m_{3n}\leq j$ such that $\sum_{i=1}^{3n}m_i=j$.
 For $n=1$ one readily recognizes the well-known eigenfunctions
\cite{Risk} for the linear Fokker--Planck operator in $\Rset^3$,
viz. r.h.s.(\ref{FPkinSIMPLE}) with constant
$\vareps_0$ and $\uV_0$, easily calculated by separation of variables.
 In fact, what we have recovered are precisely the eigenfunctions for the
linear Fokker--Planck operator in $\Rset^{3n}$, see
 (\ref{nDIFFhierarchyEQlim}).

 Now assume that one can choose sequences of initial conditions
$F^{(N)}_0$ such that, for each fixed $j$ and $\ell$, the Fourier coefficients
$F_{j,\ell}^{(N)}$ converge to a limit $F_{j,\ell}$ \emph{such that} each
initial $n$-velocity marginal density, $n\in\Nset$, converges in
$(\Lsp^2\cap\Lsp^1)(\Rset^{3n})$ to
\begin{equation}
f^{(n)}(\vV_1,\dots,\vV_n;0)
=
f_{\mathrm{M}}^{\otimes{n}}(\vV_1,...,\vV_n)
+
\sum_{j\in\Nset} \sum_{\ell\in\widetilde\Dset_j}
F_{j,\ell}
g_{j,\ell}^{(n)}(\vV_1,\dots,\vV_n);
\label{fnNULL}
\end{equation}
it then follows that the subsequent evolution of the
$n$-velocity marginal densities is given by
\begin{equation}
f^{(n)}(\vV_1,\dots,\vV_n;\tau)
=
f_{\mathrm{M}}^{\otimes{n}}(\vV_1,...,\vV_n)
+
\sum_{j\in\Nset} \sum_{\ell\in\widetilde\Dset_j}
F_{j,\ell}
g_{j,\ell}^{(n)}(\vV_1,\dots,\vV_n)
e^{- \textstyle{\frac{3j}{2\vareps_0}}\tau}.
\label{fnSOL}
\end{equation}
 Formula \refeq{fnSOL} describes an exponentially fast approach
to equilibrium in the ensemble of infinite systems.
 The $f^{(n)}(\,.\,;\tau)\in(\Lsp^2\cap\Lsp^1)(\Rset^{3n})$, and
in addition they automatically satisfy
\begin{eqnarray}
\int_{\Rset^{3n}} f^{(n)}(\vV_1,\dots,\vV_n;\tau)
\, d\vV_1\dots d\vV_n
\!&=&\!
1
\label{fnMASS}
\\
\int_{\Rset^{3n}}(\vV_1+\dots+\vV_n) f^{(n)}(\vV_1,\dots,\vV_n;\tau)
\, d\vV_1\dots d\vV_n
\!&=&\!
n\uV_0
\label{fnMOMENTUM}
\\
\int_{\Rset^{3n}} \frac12
(|\vV_1|^2+\dots +|\vV_n|^2)
f^{(n)}(\vV_1,\dots,\vV_n;\tau)\, d\vV_1\dots d\vV_n
\!&=&\!
n e_0
\label{fnENERGY}
\end{eqnarray}
for all $\tau\geq 0$ (recall that $e_0 = \vareps_0 + |\uV|_0^2/2$).
 In fact, \refeq{fnSOL} solves \refeq{nDIFFhierarchyEQlim}, which now
implies that $f^{(n)}(\,.\,;\tau)$ can also be expressed through
integration of the initial data against the $n$-fold tensor product
of \refeq{OUkernel}.
 The upshot is that $f^{(n)}(\,.\,;\tau)\in \Ssp(\Rset^{3n})\ \forall\tau>0$
(Schwartz space).
 To vindicate these conclusions, for us it remains  to show that
the infinitely many constraints on each $F_{j,\ell}$ implied by 
\refeq{fnNULL}, viz.
\begin{equation}
F_{j,\ell}
=
\frac{\langle f^{(n)}_0|g_{j,\ell}^{(n)}\rangle}
{\langle g^{(n)}_{j,\ell}|g_{j,\ell}^{(n)}\rangle}\qquad \forall n\in\Nset,
\label{FjlCONSTRAINTS}
\end{equation}
where $\langle\,.\,|\,.\,\rangle$ now
means inner product in $\Lsp^2(\Rset^{3n})$, do not impose impossible
consistency requirements.
 To show this, recall that the $f^{(n)}_0$ by definition satisfy
\begin{equation}
\int_{\Rset^3} f_{0}^{(n+1)}(\vV_1,\dots,\vV_{n+1}) d\vV_{n+1}
=
f_{0}^{(n)}(\vV_1,\dots,\vV_{n}),
\label{fMARG}
\end{equation}
which in view of  \refeq{fnNULL} implies
that the hierarchy of the $g_{j,\ell}^{(n)}$ must satisfy
\begin{equation}
\int_{\Rset^3} g_{j,\ell}^{(n+1)}(\vV_1,\dots,\vV_{n+1}) d\vV_{n+1}
=
g_{j,\ell}^{(n)}(\vV_1,\dots,\vV_{n})\prod_{i=1}^3 \delta_{k_{3(n+1)-i},0},
\label{gMARG}
\end{equation}
which is readily verified by explicit integration of
\refeq{nEIGlim}.
 Thus, the constraints \refeq{FjlCONSTRAINTS} are automatically consistent,
and this vindicates our initial assumption.


\section{Propagation of Chaos}


Setting $n=1$ in \refeq{nDIFFhierarchyEQlim}, and
changing the time scale by setting $\tau = \frac{2}{3}\vareps_0t$,
we recover  (\ref{FPkinSIMPLE}), with $f^{(1)}$ in place of $f$.
 However,  (\ref{FPkinSIMPLE}) (or \refeq{FPkin} for
that matter) cannot be said to have been shown to be a kinetic
equation yet.
 Note that propagation of chaos has not entered the
derivation of \refeq{nDIFFhierarchyEQlim}.
 In fact,  (\ref{nDIFFhierarchyEQlim}) for
$n=1,2,\dots$ constitutes a ``Fokker--Planck hierarchy'' analogous to the
the well-known Boltzmann, Landau and Vlasov hierarchies which arise in the
validation of kinetic theory \cite{SpoBOOK,CIPbook} using ensembles.
 In our case the hierarchy has the very simplifying feature
that the $n$-th equation in the hierarchy is decoupled from the
equation for the $n+1$-th marginal.
 Since all the hierarchies used in the validation of kinetic theory
are by construction
\emph{linear}\footnote{More precisely, they are only essentially linear,
        for the parameters $\vareps_0$ and $\uV_0$, which also enter
        any of the other hierarchies whenever they describe ensembles
        of systems conserving mass, momentum, and energy, are all
        tied up with the initial conditions.}
in the ``vector''
of the $f^{(n)}$, whenever one has a decoupling hierarchy one
obtains closed linear equations for the $f^{(n)}$.
 In particular, our equation  \refeq{nDIFFhierarchyEQlim}
with $n=1$ is already a closed linear equation for $f^{(1)}$.
 However, at this point, any $f^{(n)}$ is still in general
an ensemble superposition of states; in particular,
$f^{(1)}$ still describes a statistical ensemble of pure states
$f$ with same mass, momentum, and energy.
 By ignoring this fact one can mislead oneself into thinking that
\refeq{nDIFFhierarchyEQlim} with $n=1$ and $f^{(1)}$ in place of $f$
is already the kinetic equation we sought.

 The final step in extracting \refeq{FPkinSIMPLE} as kinetic equation
for the pure states involves the Hewitt--Savage \cite{HewSav}
decomposition theorem.
 This theorem says that in the continuum limit any $f^{(n)}$
is a unique convex linear superposition of extremal (i.e. pure) $n$ particle
states, and that these pure states are products of $n$ identical
one-particle functions $f$ evaluated at $n$ generally
different velocities.
   Each of the $f$ in the support of the superposition measure
represents the velocity density function of an actual individual member
of the infinite statistical ensemble of infinitely-many-particles
systems.
 In formulas, at $\tau =0$ the initial data for $f^{(n)}$ read
\begin{equation}
f^{(n)}(\vV_1,\dots,\vV_n;0)
=
\langle f_0^{\otimes{n}}(\vV_1,...,\vV_n)\rangle,
\label{HWinitially}
\end{equation}
where $\langle\,.\,\rangle$ is the Hewitt--Savage \cite{HewSav}
ensemble decomposition measure on the space of initial velocity
density functions $f_0$ of {individual physical systems}
with same mass $m(f_0)(=1)$, momentum $\pV(f_0) = \uV_0$ and
energy $e(f_0) = e_0 = \vareps_0 + |\uV_0|^2/2$.
 To extend this representation to $\tau>0$, let $U_\tau^{(n)}$ denote the
one-parameter evolution semigroup for \refeq{nDIFFhierarchyEQlim}, i.e.
$f^{(n)}(\vV_1,\dots,\vV_n;\tau)=U_\tau^{(n)}f^{(n)}_0(\vV_1,\dots,\vV_n)$.
 Noting now that the Hewitt--Savage measure is of course invariant
under the evolution, and that by the linearity of \refeq{nDIFFhierarchyEQlim}
it commutes with the linear operator $U_\tau^{(n)}$ for all $\tau\geq 0$, it
follows that at later times $\tau >0$ the $n$ point density of the ensemble
is given by
\begin{equation}
f^{(n)}(\vV_1,\dots,\vV_n;\tau)
=
\langle U_\tau^{(n)}f^{\otimes{n}}_0(\vV_1,...,\vV_n)\rangle.
\end{equation}
 This so far simply states that, if the ensemble is initially a statistical
mixture of pure states (product states), then at later times it is a
statistical mixture of time-evolved  initially pure states.
 Next we note that by inspection of \refeq{nDIFFhierarchyEQlim} it follows
that
\begin{equation}
U_\tau^{(n)}f^{\otimes{n}}_0(\vV_1,...,\vV_n)
=
(U_\tau^{(1)}f_0)^{\otimes{n}}(\vV_1,...,\vV_n),
\end{equation}
viz. pure states evolve into pure states.
 Every factor $f(\vV_k;\tau)= U_\tau^{(1)}f_0(\vV_k)$
solves  \refeq{FPkinSIMPLE} with $\tau = \frac{2\vareps_0}{3}t$, obeying
the desired conservation laws.
 At last one can legitimately say that  (\ref{FPkinSIMPLE})
has been derived as a full-fledged kinetic equation valid for almost
every (w.r.t. $\langle\,.\,\rangle$) individual member of the limiting
ensemble.


\section{Summary and Outlook}


  In summary, the diffusion equation on $\Mset^{3N-4}_{\uV_0,e_0}$ can be 
interpreted as the simplest ``master equation'' for an underlying 
$N$-body Markov process with single-particle and pair terms.
 The $N\to\infty$ limit for the marginal densities of solutions to 
the diffusion equation is well-defined and can be carried out explicitly.
 After invoking the Hewitt--Savage decomposition, the limit $N\to\infty$
is seen to produce solutions of the ``kinetic Fokker--Planck equation''
describing individual isolated systems conserving  mass, momentum, and energy.
 The Fokker--Planck equation \refeq{FPkin} is exactly solvable and  
displays correctly the qualitative behavior of a typical kinetic 
equation. 
 In this sense,  (\ref{FPkin}) really can be regarded
as the simplest example of a kinetic equation
of the ``diffusive" type, in the same family as, for instance, the much
more complex Landau and Balescu-Lenard-Guernsey equations.

 Our work raises many new questions.
1) In particular, in Appendix Ab we have only written down the generator 
for the adjoint process of the underlying $N$-particle Markov process; 
hence, what is the explicit characterization of this process?
2) A derivation of a kinetic equation \`a la Kac is an intermediate step 
towards a full validation from some deterministic (Hamiltonian) microscopic 
model, which is in general a very difficult program, see the rigorous 
derivations of kinetic equations in \cite{SpoBOOK,CIPbook}.  
 The substitute Markov process is usually chosen to preserve some of the 
essential features of the deterministic dynamics which (formally) leads 
to the same kinetic equation. 
 Here we have only identified a stochastic model which leads to \refeq{FPkin}.
 Villani's work \cite{Vil98} suggests that a deterministic model may exist 
which in the kinetic regime leads to \refeq{FPkin}.
 Can one indentify this model?
3)  In this paper, we conveniently assumed that the Fourier coefficients
ensure convergence of the marginal density functions in $\Lsp^2\cap\Lsp^1$
and subsequently upgraded the regularity to Schwartz functions. 
 What are the explicit conditions on the Fourier coefficients
of the initial functions on $\Mset^{3N-4}_{\uV_0,e_0}$ which ensure
convergence  in $\Lsp^2\cap\Lsp^1$, in Schwartz space, in some topology 
for measures?
4)
 Since the PDEs in our finite-$N$ Fokker--Planck hierarchy 
are already self-contained for each $n$ (viz., they do not involve 
the usual coupling to $f^{(n+1)}$), the finite-$N$ corrections to the 
limiting evolutions can be studied in great detail; hence, for instance, 
how do the explicit corrections to propagation of chaos look? 
5) We already mentioned in a footnote that the kinetic Fokker--Planck
equation can easily be generalized to situations where the system is
exposed to some external driving force by adding a forcing term.
 Can one derive this equation from some suitable ensemble of driven systems?
 Under which conditions do there exist stationary
non-equilibrium states, and what are their stability properties?
6) Finally, our derivation is only valid for the space-homogeneous
Fokker--Planck equation without driving force term; hence, can one 
extend our derivation to obtain the space-inhomogeneous generalization 
of the kinetic Fokker--Planck equation, first without and then with
driving force term?
 These are many interesting questions which should be answered in 
future works.
\medskip

\noindent
\textbf{Acknowledgment} 
We thank the referees for drawing our attention to \cite{BakMaz} and
for their constructive criticisms. Thanks go also to Michael Loss
for pointing out Mehler's paper \cite{Mehler}.
Kiessling was supported by NSF Grant DMS-0103808. 
Lancellotti was supported by NSF Grant DMS-0318532.

\newpage

\section*{Appendix}

\subsection*{A. Two useful representations of the Laplacian on spheres}

\subsubsection*{Aa. Extrinsic representation in divergence form}

  For the purpose of obtaining equations for the marginals
by integrating  (\ref{heat}), 
it is advantageous to express the Laplacian on the right-hand side
in terms of the projection operator 
$P_{\Mset^{3N-4}_{\uV_0,e_0}}$ from $\Rset^{3N}$ 
to the fibers of the tangent bundle of the embedded manifold  
$\Mset^{3N-4}_{\uV_0,e_0}$.  It is easy to verify \cite{KieLan04} that
\begin{equation}
\Delta_{\Mset^{3N-4}_{\uV_0,e_0}} F^{(N)}
=
\nabla\cdot[P_{\Mset^{3N-4}_{\uV_0,e_0}}\nabla F^{(N)}]
\label{LapBel2}
\end{equation}
 In order to have an explicit expression for 
$P_{\Mset^{3N-4}_{\uV_0,e_0}}$
we introduce an orthogonal basis for the orthogonal complement
of the tangent space to $\Mset^{3N-4}_{\uV_0,e_0}$
at $\vVN\in\Mset^{3N-4}_{\uV_0,e_0}\subset \Rset^{3N}$.
 Clearly, such orthogonal complement is spanned by the four vectors $\vVN$ and
$\eVN_\sigma=(\eV_{\sigma},\dots,\eV_{\sigma})$, $\sigma=1,2,3$,
where the $\eV_{\sigma}$ are the standard unit vectors in $\Rset^3$. 
 The vectors $\eVN_\sigma$ are orthogonal to each other but not to
$\vVN$; projecting away the non-orthogonal component of $\vVN$ yields
\begin{equation}
\biggl(
\id_{3N} -\frac{1}{N}\sum_{\sigma=1}^3 \eVN_\sigma\otimes \eVN_\sigma
\biggr)
\cdot\vVN
=
\vVN - \uVN.
\end{equation}
 The vectors $\{\vVN-\uVN, \eVN_1, \eVN_2, \eVN_3\}$ form the desired
orthogonal basis; their magnitudes are $\abs{\eVN_\sigma}=\sqrt{N}$ and
$\abs{\vVN-\uVN}=\sqrt{2N\vareps_0}$.
 Finally,  (\ref{LapBel2}) becomes
\begin{equation}
\!\!\!\!\Delta_{\Mset^{3N-4}_{\uV_0,e_0}} F^{(N)} 
=
{\partial_\vVN}\cdot
\!\left[\!\left(\!\id_{3N}-
\frac{1}{N} \sum_{\sigma=1}^3\eVN_\sigma\otimes\eVN_\sigma-
\frac{1}{2N\vareps_0}
(\vVN\!-\!\uVN)\otimes(\vVN\!-\!\uVN)\!\right)\!{\partial_\vVN F^{(N)}}
\!\right]\!
\label{heat1}
\end{equation}

\subsubsection*{Ab. Representation for the $N$-Body Markov Process}

In the main part of this paper we started from the diffusion equation on
the manifold $\Mset^{3N-4}_{\uV_0,e_0}$ of $N$-body systems with same 
energy (per particle) $e_0$ and momentum (per particle) $\uV_0$, then
took the limit $N\to\infty$, obtaining the kinetic Fokker--Planck
equation \refeq{FPkinSIMPLE}, which rewrites into \refeq{FPkin}
in view of the conservation laws. 
 The Laplace--Beltrami operator on $\Mset^{3N-4}_{\uV_0,e_0}$
is the generator of the adjoint semigroup of the 
underlying stochastic Markov process that rules the microscopic 
dynamics of an individual $N$-body system.
  Here we show that this generator can be written as a 
sum of single particle and two-particle operators, thus characterizing
the Markov process as a mixture of individual stochastic motions and
stochastic binary interactions. 
 Moreover, we show that the binary
particle operators are the only ones that do not vanish in the $N\to\infty$
limit.  
  This means that the kinetic Fokker--Planck equation can also
be derived in terms of an $N$-body stochastic process with
purely binary interactions, which is more satisfactory
from a physical point of view.

 Recall that in section 2 we explained that 
$\Mset^{3N-4}_{\uV_0,e_0}$ can be identified with the sphere 
$\Sset^{3N-4}_{\sqrt{2N\vareps_0}}$ centered at the origin
of $\Lset^{3N-3}$ (which itself is an affine linear subspace
of the space of all velocities, $\Rset^{3N}$).
 Recall that 
$\Delta_{\Sset^{3N-4}_{\sqrt{2N\vareps_0}}} 
 = \frac{1}{2N\vareps_0}\Delta_{\Sset^{3N-4}}$.
  Note the well-known representation
\begin{equation}
\Delta_{\Sset^{3N-4}}
=
\!\!\!\!\!\!\!\!{\sum_{\qquad 1\leq k < l\leq 3(N-1)}}\!\!
\Big(
w_{k} \partial_{w_{l}} -w_{l} \partial_{w_{k}}
\Big)^2,
\label{LAPopDECOMPOSED}
\end{equation}
where ${w_{k}}$ is the $k$-th Cartesian component of 
$\wVN\in\Sset^{3N-4}\subset\Rset^{3(N-1)}$
(note that in section 2 we used $\wVN\in\Sset^{3N-4}_{\sqrt{2N\vareps_0}}$,
but note furthermore that the r.h.s. of \refeq{LAPopDECOMPOSED} is invariant
under $\wVN\to\lambda\wVN$).
 Grouping the components of $\wVN$ into blocks of vectors $\wV_k\in\Rset^3$,
$k=1,...,N-1$, the r.h.s. of \refeq{LAPopDECOMPOSED} can be recast as
\begin{eqnarray}
\Delta_{\Sset^{3N-4}} 
= 
&&\!\!\!\!
\Big. \sum_{k=1}^{N-1}
\sum_{\stackrel{l=1}{l\ne k}}^{N-1}
\Big(3\wVk\cdot\pdwk+\abs{\wVk}^2\pdwl\cdot\pdwl 
-
\big(
\wVk\cdot\pdwk
\big)
\big(\wVl\cdot\pdwl
\big)
\Big)
\nonumber\\
&&- 
\sum_{k=1}^{N-1} 
\big(
\wVk\times\pdwk
\big)^2
\Big.
,
\label{MASTERopDECOMPOSED}
\end{eqnarray}
containing one-body terms as well as binary terms.
 Note however that the first term in the binary sum is 
effectively a sum of two-body terms in disguise, which
scale with factor $N-2$ and thus survive in the limit $N\to\infty$, 
while the true one-body sum (second line) drops out in that limit. 
 This implies that the kinetic Fokker--Planck equation \refeq{FPkinSIMPLE}
can be derived from a master equation on $\Mset^{3N-4}_{\uV_0,e_0}$ 
which contains \emph{only} the binary terms (first line)
in \refeq{MASTERopDECOMPOSED}.
 This in turn implies that \refeq{FPkinSIMPLE} is the kinetic equation
for an underlying system of $N$ particles with stochastic pair interactions.

\subsection*{B. High-Dimension Asymptotics of Associated Legendre Functions}

 In  \refeq{eigmarg2} the associated Legendre functions of degree 
$s=0,1,2,...$ and order $r=0,...,s$ in $q$ dimensions occur.
 They are defined on the interval $[-1,1]$ and given by 
\begin{equation}
\widetilde{P}_s^r(t;q)
=
\sqrt{q}^{s+r}
\frac{s!}{2^r}\,\Gamma\left(\frac{q-1}{2}\right)
\sum_{l=0}^{\intgpart}
\left(-\frac{1}{4}\right)^l
\frac{(1-t^2)^{l+\frac{r}{2}}\, t^{s-r-2l}}
{l!\,(s-r-2l)!\,\Gamma\left(l+r+\frac{q-1}{2}\right)}
\label{LEGENDREf}
\end{equation}
which differ from the $P_s^r(t;q)$ in \cite{Mul} in their normalization.
 In our investigation, $q = 3N - p$ and $t = \frac{w}{\smallr}$,
and we are interested in the limit $N\to\infty$.

 The familiar asymptotics of Euler's Gamma function gives us
\begin{equation}
\frac{\Gamma\left(x\right)}
     {\Gamma\left(a+x\right)}
=
x^{-a} + O\left(x^{-(a+1)}\right).
\end{equation}
for $x\gg 1$. 
 Applying this asymptotics with $2x = q-1=3N-p-1$ and $a=l+r$ to
\refeq{LEGENDREf}, 
we find that given $p\in\Nset$ and $w\in \Rset$ (which implies
$N> \max\{p/3,w^2/(2\vareps_0)\}$), when $N\gg 1$ we have
\begin{eqnarray}
\widetilde{P}_s^r\left(\frac{w}{\smallr};3N\!-\!p\right)
=&&\!\!
\sqrt{2}^{r-s}
\sum_{l=0}^{\intgpart}
(-1)^l \frac{s!}{l!\,(s-r-2l)!}
\left(\sqrt{{\textstyle{\frac{3}{\vareps_0}}}}\,w\right)^{s-r-2l}
\nonumber
\\
&&\!\!
+\; O\!\left(\frac{1}{\sqrt{N}} \right).
\end{eqnarray}
  By comparing with the formula for the Hermite polynomial of degree $k$
on $\Rset$, 
\begin{equation}
H_{k}(x)
=
\sum_{l=0}^{\big\lfloor\!\frac{{\scriptstyle k}}{2}\!\big\rfloor}
(-1)^{l+k}\frac{ s!}{l!\,(k-2l)!}(2x)^{k-2l},
\end{equation}
we see that, given $p\in\Nset$ and $w\in \Rset$, we have
\begin{eqnarray}
\widetilde{P}_s^r\left(\frac{w}{\smallr};3N-p\right)
=
\left(-\sqrt{2}\right)^{r-s}\, 
H_{s-r}\left(\sqrt{{\textstyle{\frac{3}{4\vareps_0}}}}w\right)
+\; O\!\left(\frac{1}{\sqrt{N}} \right)
\end{eqnarray}
when $N\gg 1$.
 Hence, for all fixed $p$ we now find that pointwise for any $w\in\Rset$,
\begin{equation}
\lim_{N\to\infty}
\widetilde{P}_s^r\left(\frac{w}{\smallr};3N-p\right)
=
\left(-\sqrt{2}\right)^{r-s}\,
H_{s-r}\left(\sqrt{{\textstyle{\frac{3}{4\vareps_0}}}}w\right)
\end{equation}
where again it is understood that
$N> \max\{p/3,w^2/(2\vareps_0)\}$ in the expression under the 
limit in the left-hand side.
 Equation (\ref{eigmarg3}) in the main text follows.


\end{document}